\documentclass[a4paper, pra, twocolumn,superscriptaddress,showpacs]{revtex4}

\usepackage{amssymb}
\usepackage{amsmath}
\usepackage{epsfig}
\usepackage{color}
\usepackage{graphics, graphicx}
\usepackage{bbold}
\usepackage{psfrag}
\usepackage{mathcomp}
\usepackage{subfigure}
\usepackage{verbatim}
\usepackage{color}
\usepackage[colorlinks,citecolor=blue]{hyperref}

\begin{document}

\date{\today}

\title{Swallowtail Structure in Fermi Superfluids with Periodically Modulated Interactions}

\author{Dongyang Yu}
\affiliation{Department of Physics, Renmin University of China, Beijing 100872, China}
\author{Wei Yi}
\email{wyiz@ustc.edu.cn}
\affiliation{Key Laboratory of Quantum Information, University of Science and Technology of China, Chinese Academy of Sciences, Hefei, Anhui, 230026, China}
\affiliation{Synergetic Innovation Center of Quantum Information and Quantum Physics, University of Science and Technology of China, Hefei, Anhui 230026, China}
\author{Wei Zhang}
\email{wzhangl@ruc.edu.cn}
\affiliation{Department of Physics, Renmin University of China, Beijing 100872, China}
\affiliation{Beijing Key Laboratory of Opto-electronic Functional Materials and Micro-nano Devices,
Renmin University of China, Beijing 100872, China}

\begin{abstract}
We study the superfluid flow in a quasi-one-dimensional Fermi gas with spatially modulated interactions induced by an optical Feshbach resonance. Due to the competition between the periodicity of the modulated interaction and the nonlinearity of the background interaction, an interesting swallowtail structure emerges in the energy spectrum under appropriate parameters. As the interaction strengths are tuned, the swallowtail structure may disappear, giving rise to various different states on a rich phase diagram. We investigate the spatial distribution of the order parameter and the particle density under various parameters, which are useful for the experimental detection of the interesting phases in these systems.
\end{abstract}
\pacs{67.85.Lm, 03.75.Ss, 05.30.Fk}

\maketitle

\section{introduction}
\label{sec:intro}
One of the central topics in the research of cold atoms is the study of peculiar properties of Bose and Fermi superfluids.
Along this line, much effort has been devoted to unveil various exotic features of
superfluidity, e.g., collective excitations~\cite{jin-96, mewes-96, bartenstein-04}, vortex lattice structure~\cite{abo-shaeer-01, zwierlein-05}, the propagation of sound~\cite{andrews-97, joseph-07, sidorenkov-13}, and the generation of solitons~\cite{denschlag-00}. Among them, an interesting phenomenon is the presence of loop structures in the energy dispersion of a superfluid flowing within a periodic potential.
As first discussed in the context of Bose-Einstein condensate (BEC)~\cite{wu-02, diakonov-02, zhang-13}, this loop,
or the so-called swallowtail structure, is a direct consequence of the competition between
the nonlinear interaction and the periodic potential. While the periodic potential leads to a spatially modulated atom number density and a band structure of Bloch states, the repulsive interaction between atoms favors a uniform distribution and tends to retain the quadratic dispersion. As a result, the swallowtail structure appears at the boundaries of the Brillouin zone for strong enough interactions~\cite{mueller-02}.

For Fermi systems, the existence of swallowtail structures has been discussed recently for a two-component
Fermi gas flowing in a one dimensional (1D) optical lattice~\cite{watanabe-11}. By tuning the interaction
between the two spin states from the weakly interacting Bardeen-Cooper-Schrieffer (BCS) limit to the
BEC limit via a magnetically tuned Feshbach resonance, a swallowtail structure can emerge in the energy dispersion $E_S(Q)$ of a supercurrent flowing along the lattice with a wave vector $Q$. Although swallowtail structures in Fermi systems reflect different properties of superfluidity as compared to the Bose counterpart~\cite{watanabe-11}, the physics underlying the phenomenon is similar, i.e., the competition between the nonlinear interactions and the band structure induced by the periodical external potential.

Besides imposing an external lattice potential, another route to implementing a periodical environment is via an optical Feshbach resonance~\cite{fedichev-96, theis-04,fu-13}. In such a process, the coupling between the open and the closed channels of the Feshbach resonance is driven by a laser beam, so that the effective interaction between the atoms depend on the intensity of the laser field. If the optical Feshbach resonance is driven by a standing wave, the resulting interaction will naturally acquire a spatially modulated component with the same period as that of the standing wave. Considering the recent experimental progress on optical Feshbach resonances, it is of great interest to investigate the effects of this spatially modulated interaction on the properties of a Fermi superfluid, and particularly, on the existence and characteristics of a swallowtail structure.

In this paper, we study the properties of superfluid phases in a quasi-1D two-component
Fermi gas with an interspecies interaction which is spatially modulated along the elongated ($x$) direction.
The effective 1D interaction contains a uniform component, which is assumed to be attractive $U_1 <0$;
and a spatially modulated portion $U_2 \cos (2 k_{\rm ofr} x)$, with $k_{\rm ofr}$ the wave vector of the laser
used to drive the optical Feshbach resonance. We adopt a mean-field approach and solve the real-space Bogoliubov-de Gennes (BdG) equations
for pairing states with different supercurrent wave vector $Q$. We find that, given a strong enough uniform interaction $|U_1|$ and a weak enough modulated interaction $U_2$, a swallowtail structure can emerge in the energy dispersion $E_S(Q)$ for $Q$ near $|k_{\rm ofr}/2|$, i.e., near the boundaries of the first Brillouin zone. This is consistent with the previous understanding that the swallowtail structure is a signature of dominating nonlinear interactions over the periodicity of the system. In the opposite limit, i.e., with small $|U_1|$ and large $U_2$, the ground state of the system becomes a Fulde-Ferrell-Larkin-Ovchinnikov (FFLO)-like superfluid with a finite center-of-mass momentum $Q=k_{\rm ofr}/2$~\cite{ff, lo}. In the intermediate region between these two limiting cases, the ground state is either a superfluid with a critical velocity characterized by the critical wave vector $Q_c < k_{\rm ofr}$, or a superfluid which is robust for arbitrary $Q$ within the first Brillouin zone. We map out the phase diagram on the $|U_1|$-$U_2$ plane and demonstrate the spatial distributions of order parameter and number distribution, which are helpful for the experimental detection of the different phases.

The remainder of the manuscript is organized as follows. In Sec.~\ref{sec:formalism}, we introduce the model system and the mean-field approach we have adopted with a brief discussion on numerical algorithms. We discuss in detail the swallowtail structure in Sec.~\ref{sec:sw}, and map out the phase diagram in Sec.~\ref{sec:pd}. In Sec.~\ref{sec:pd}, we also characterize the FFLO-like state identified on the phase diagram. Finally, we summarize in Sec.~\ref{sec:summary}.


\section{Formalism}
\label{sec:formalism}

We consider a two-component Fermi gas loaded in a quasi-1D confinement, where the Hamiltonian can be written as
\begin{eqnarray}
H&=&\int d{\bf r} \sum_{\sigma} \Psi^\dagger_{\sigma}({\bf r})
\left[ -\frac{\hbar^2\nabla_{\bf r}^2}{2m}+ \frac{1}{2}m \omega_\perp^2 (y^2 + z^2)\right] \Psi_{\sigma}({\bf r})
\nonumber \\
&&+\int d{\bf r} g({\bf r}) \Psi^\dagger_{\uparrow}({\bf r}) \Psi^\dagger_{\downarrow}({\bf r})
\Psi_{\downarrow}({\bf r}) \Psi_{\uparrow}({\bf r}).
\end{eqnarray}
Here, $\Psi_\sigma({\bf r})$ ($\Psi_\sigma^\dagger({\bf r})$) is the annihilation (creation) operator for fermions with pseudo-spin (hyperfine state) $\sigma = (\uparrow, \downarrow)$ and mass $m$ at position ${\bf r}$. The bare interaction $g$ is related to the three-dimensional $s$-wave scattering length $a_s$ via the renormalization relation $1/g = m/(4 \pi \hbar^2 a_s) - \sum_{\bf k} 1/2 \epsilon_{\bf k}$, with the single-particle dispersion $\epsilon_{\bf k} = \hbar^2 k^2/(2m)$. For an optical Feshbach resonance, the scattering length $a_s({\bf r})$ can be position dependent if the intensity of the tuning laser acquires a spatial modulation.

In the quasi-1D limit where the transversal confinement frequency $\omega_\perp$ is much larger than other energy scales in the system, the motion of an atom within the radial plane will be frozen into the ground harmonic state if it is well separated from other atoms. This allows us to integrate out the transversal degrees of freedom and rewrite the Hamiltonian into an effective 1D form
\begin{eqnarray}
\label{eqn:H}
H_{\rm eff}&=&\int dx  \sum_{\sigma} \psi^\dagger_{\sigma}(x)
\left( -\frac{\hbar^2 \partial_x^2}{2m} \right) \psi_{\sigma}(x) \nonumber \\
&&+ \int dx g_{\rm 1D}(x) \psi^\dagger_{\uparrow}(x) \psi^\dagger_{\downarrow}(x)
\psi_{\downarrow}(x) \psi_{\uparrow}(x),
\end{eqnarray}
where $\psi_\sigma(x)$ and $\psi_\sigma(x)$ are the 1D fermionic operators. The 1D effective interaction can be related to the $s$-wave scattering length via the confinement-induced resonance~\cite{olshanii-98}
\begin{equation}
\label{eqn:g1d}
g_{\rm 1D} = \frac{4 \pi \hbar^2 a_s}{m} \frac{1}{\pi a_\perp^2} \left(1 - {\cal C} \frac{a_s}{a_\perp} \right)^{-1},
\end{equation}
where $a_\perp \equiv \sqrt{\hbar/m \omega_\perp}$ is the characteristic length of the transversal harmonic trapping potential, and ${\cal C} = \lim_{s\to \infty} (\int_0^\infty ds^\prime/\sqrt{s^\prime} - \sum_{s^\prime=1}^s 1/\sqrt{s^\prime}) \approx 1.4603$.

We then consider the case where the scattering length is tuned by a standing wave along the $x$ direction, via an optical Feshbach resonance between the open-channel threshold and a molecular state in the closed channel. Since the molecular level is only a metastable state with a finite decay rate $\Gamma$ due to spontaneous emission, the $s$-wave scattering length is typically complex
\begin{eqnarray}
\label{eqn:comp-as}
{\tilde a}_s = a_{\rm bg} \left[ 1 + \frac{W_0(I)}{\hbar(\omega_\ell - \omega_c) + i\Gamma/2} \right].
\end{eqnarray}
Here $a_{\rm bg}$ is the background scattering length, the resonance width $W_0 \equiv \gamma_0 I(x)$ is proportional to the laser intensity $I$, $\omega_\ell$ is the laser frequency, and $\omega_c$ characterizes the detuning between the open and the closed channels of the Feshbach resonance.
Under the practical condition of $W_0 \gg \Gamma$, the real part of Eq.~(\ref{eqn:comp-as}) dominates its imaginary part, and the scattering length takes an approximate form in the vicinity of the resonance
\begin{eqnarray}
\label{eqn:real-as}
{\rm Re}[{\tilde a}_s] &\approx& a_{\rm bg} + a_{\rm bg} \gamma_0 I \frac{\hbar (\omega_\ell - \omega_c)}{\hbar^2 (\omega_\ell - \omega_c)^2 + \Gamma^2/4}.
\end{eqnarray}
This result thus leads to a spatially modulated $s$-wave scattering length $a_s(x) \equiv {\rm Re}[{\tilde a}_s]$ via the position-dependent laser intensity $I(x) = I_0 \cos^2(k_{\rm ofr} x)$, where $k_{\rm ofr}$ is the wave vector. Substituting Eq. (\ref{eqn:real-as}) into (\ref{eqn:g1d}), and focusing on the regime away from the confinement-induced resonance with $a_s \ll a_\perp$, we obtain an effective 1D interaction with spatial modulations~\cite{qi-11}
\begin{eqnarray}
\label{eqn:g1d-2}
g_{\rm 1D}(x) &=& U_1 + U_2 \cos(2 k_{\rm ofr} x),
\end{eqnarray}
where
\begin{eqnarray}
\label{eqn:U1U2}
U_1 &=& \frac{4\pi \hbar^2 a_{\rm bg}}{m} \frac{1}{\pi a_\perp^2}
\left[1+ \frac{\gamma_0}{2} \frac{\hbar (\omega_\ell - \omega_c)}{\hbar^2 (\omega_\ell - \omega_c)^2 + \Gamma^2/4}\right],
\nonumber \\
U_2 &=&  \frac{4\pi \hbar^2 a_{\rm bg}}{m} \frac{\gamma_0}{2 \pi a_\perp^2} \frac{\hbar (\omega_\ell - \omega_c)}{\hbar^2 (\omega_\ell - \omega_c)^2 + \Gamma^2/4}.
\end{eqnarray}

We solve the Hamiltonian Eq. (\ref{eqn:H}) via the mean-field approach by defining pairing order parameter
$\Delta(x) \equiv g_{\rm 1D}(x) \langle \psi_{\downarrow}(x) \psi_{\uparrow}(x) \rangle$. The resulting Hamiltonian thus takes the following form
\begin{eqnarray}
\label{eqn:H-mf}
H_{\rm MF} &=& \int dx \Bigg\{ \sum_{\sigma} \psi^\dagger_{\sigma}\left(-\frac{\hbar^2\partial^2_x}{2m}+\bar{V}(x) - \mu \right)\psi_{\sigma}
\nonumber \\
&&
+\left( \Delta(x)\psi^\dagger_{\uparrow}\psi^\dagger_{\downarrow}+ \Delta^*(x)\psi_{\downarrow} \psi_{\uparrow}\right)
\nonumber \\
&&
- g_{\rm 1D}(x) n^2(x) - \frac{|\Delta(x)|^2}{g_{\rm 1D}(x)}\Bigg\},
\end{eqnarray}
where the chemical potential $\mu$ is assumed to be identical for both spin species,
the Hartree term $\bar{V}(x) \equiv g_{\rm 1D}(x) \langle \Psi^\dagger_{\uparrow}(x)\Psi_{\uparrow}(x) \rangle
=g_{\rm 1D}(x) n(x)$ with $n(x) = n_\uparrow(x) = n_\downarrow(x)$ representing the spatial distribution of fermions.
This mean-field Hamiltonian can be diagonalized with the Bogoliubov transformation
\begin{eqnarray}
\label{eqn:bog}
\psi_{\sigma}(x)=\sum_{\eta}u_{\eta\sigma}(x)c_{\eta\sigma} - \alpha v^*_{\eta\sigma}(x)c^\dagger_{\eta\sigma},
\end{eqnarray}
which leads to the BdG equations
\begin{eqnarray}
\label{eqn:bdg-eq1}
\left[
  \begin{array}{cc}
    K(x) & \Delta(x) \\
    \Delta^*(x) & -K(x) \\
  \end{array}
\right]\left[
         \begin{array}{c}
           u_{\eta\sigma}(x) \\
           v_{\eta\sigma}(x) \\
         \end{array}
       \right]=E_{\eta\sigma}\left[
         \begin{array}{c}
           u_{\eta\sigma}(x) \\
           v_{\eta\sigma}(x) \\
         \end{array}
       \right].
\end{eqnarray}
Here, $c^{\dagger}_{\eta,\sigma}$ ($c_{\eta,\sigma}$) is the creation (annihilation) operator of the quasiparticle with eigenenergy $E_{\eta \sigma}$, and $\alpha$ takes the value of  $+1$ ($-1$) for $\sigma=\uparrow$ ($\sigma=\downarrow$). The diagonal term reads
\begin{equation}
\label{eqn:kinetic}
K(x)=-\frac{\hbar^2\partial^2_x}{2m}-\mu+{\bar V}(x).
\end{equation}
In this paper, we focus on the case of an equal spin population, and drop the subscript of $\sigma$ in the solutions of
Eq. (\ref{eqn:bdg-eq1}) unless otherwise specified.
\begin{figure}
\includegraphics[width=8.5cm]{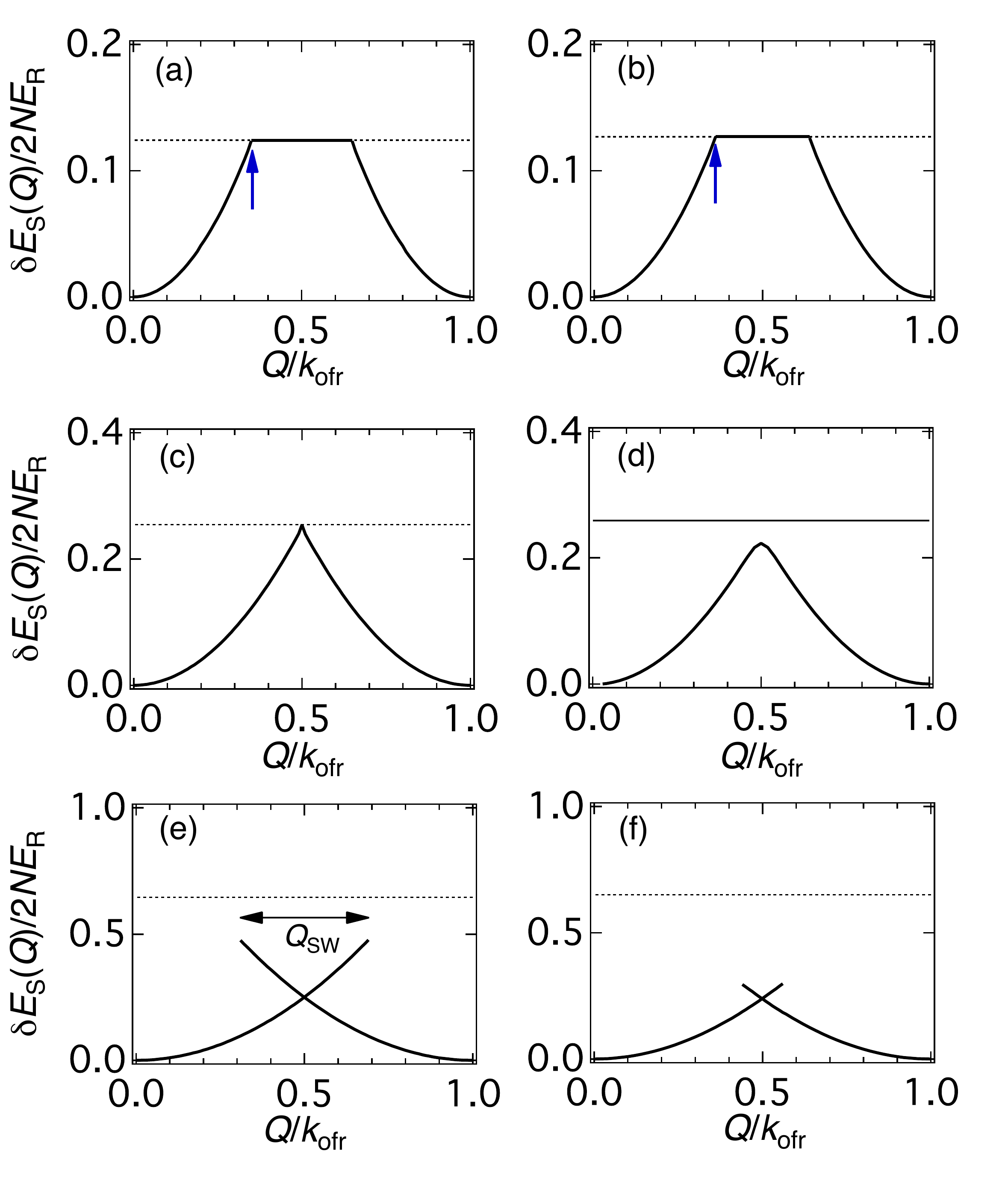}
\caption{(Color online) Energy spectra of superfluid state (solid lines) with
supercurrent wave vector $Q$ for (a)(c)(e) $U_2 n_0/E_R =0.1$ and (b)(d)(f) $U_2n_0/E_R = 0.5$.
The uniform interaction $|U_1|$ increases from top to bottom: (a)(b) $|U_1| n_0/E_R =2$,
(c)(d)  $|U_1| n_0/E_R =2.35$, and (e)(f) $|U_1| n_0/E_R =3$, .
The normal state energy $E_N$ for each parameter set is shown
as a horizontal dotted line. All energies plotted in this figure are shifted relative to the
superfluid state energy at rest $E_S(Q)$ in the corresponding cases.
Notice the emergence of the swallowtail structures as the uniform
interaction $|U_1|$ is strong enough. For weak uniform interaction $|U_1|$, the normal state
energy is smaller than the superfluid state energy at $Q = k_{\rm ofr}/2$,
leading to a critical supercurrent wave vector as indicated by arrows in (a) and (b). }
\label{fig:swallowtail-1}
\end{figure}

The pairing order parameter and the density distribution can be obtained in terms of quasiparticles via the following
expressions
\begin{eqnarray}
\label{eqn:delta-n-1}
\Delta(x)&=&g_{\rm 1D}(x) \sum_{\eta} u_\eta(x)v^*_\eta(x) f (E_\eta),\nonumber \\
n(x)&=&\sum_\eta |u_\eta(x)|^2 f (E_\eta).
\end{eqnarray}
Notice that the summation runs over all eigenstates of Eq. (\ref{eqn:bdg-eq1}), and the function
$f(s) \equiv 1/(e^{\beta s}-1)$ denotes the Fermi distribution with $\beta = 1/k_B T$ the inverse temperature.
By considering a supercurrent flowing along the $x$-direction with a wave vector $Q$, the quasiparticle wavefunctions
can be decomposed in terms of Bloch waves
\begin{eqnarray}
\label{eqn:Bloch-wave}
u_\eta(x)&=& e^{iQx} \sum_{n, k}A_{n, k,\eta}\phi_n(x)e^{ikx},\nonumber \\
v_\eta(x)&=& e^{-iQx} \sum_{n, k}B_{n, k,\eta}\phi_n(x)e^{ikx},
\end{eqnarray}
where $\phi_n(x) = e^{i 2 n k_{\rm ofr} x}$ is the wave function of the $n$-th band, and the summation of the quasimomentum $k$ runs over the first Brillouin zone. Denoting the total number of lattice sites as $N_L$, we substitute Eq. (\ref{eqn:Bloch-wave}) into (\ref{eqn:delta-n-1}), and solve Eq. (\ref{eqn:bdg-eq1}) under the particle-number constraint
\begin{eqnarray}
\label{eqn:totalnum}
N = N_\uparrow = N_{\downarrow} = N_L \int_0^{\pi / k_{\rm ofr}} n(x) dx,
\end{eqnarray}
to get the spectrum of the quasiparticles and the total energy of
the system in the superfluid phase
\begin{eqnarray}
\label{eqn:totaleng}
E_{S}= \sum_{\eta} E_{\eta} f(E_\eta).
\end{eqnarray}
This energy is then compared with the normal-state energy $E_N$, obtained by setting $\Delta(x) = 0$, to determine
the true ground state of the system.
Numerically, we solve the full problem for the lowest $N_c$ bands, and employ the plane-wave
approximation for higher bands~\cite{liu-07}. This method leads to a fairly fast convergence, and the results are
independent of the cutoff for $N_c \gtrsim 40$. In the remainder of this paper, we focus on the case of $E_F/E_R = 2.5$,
where the recoil energy $E_R \equiv \hbar^2 k_{\rm ofr}^2/2m$ is chosen as the energy unit.

\section{The Swallowtail Structure}
\label{sec:sw}

By numerically solving the BdG equation (\ref{eqn:bdg-eq1}) with a fixed filling factor $n_0 = N/N_L$,
we obtain the energy spectra of superfluid states as functions of the supercurrent wave vector $Q$ for
different interaction parameters. As shown in Fig.~\ref{fig:swallowtail-1}, the energies are relative to the
value at rest, i.e., $\delta E_S(Q) \equiv E_S(Q) - E_S(0)$.
As the parameters are tuned, an interesting swallowtail structure emerges in a regime with a weak modulated interaction $U_2$ and a strong uniform interaction $|U_1|$. As $|U_1|$ decreases, the swallowtail structure becomes less prominent and eventually disappears. These results are consistent with the general understanding that the swallowtail structure is a result of dominating nonlinear interactions.
\begin{figure}
\includegraphics[width=8.5cm]{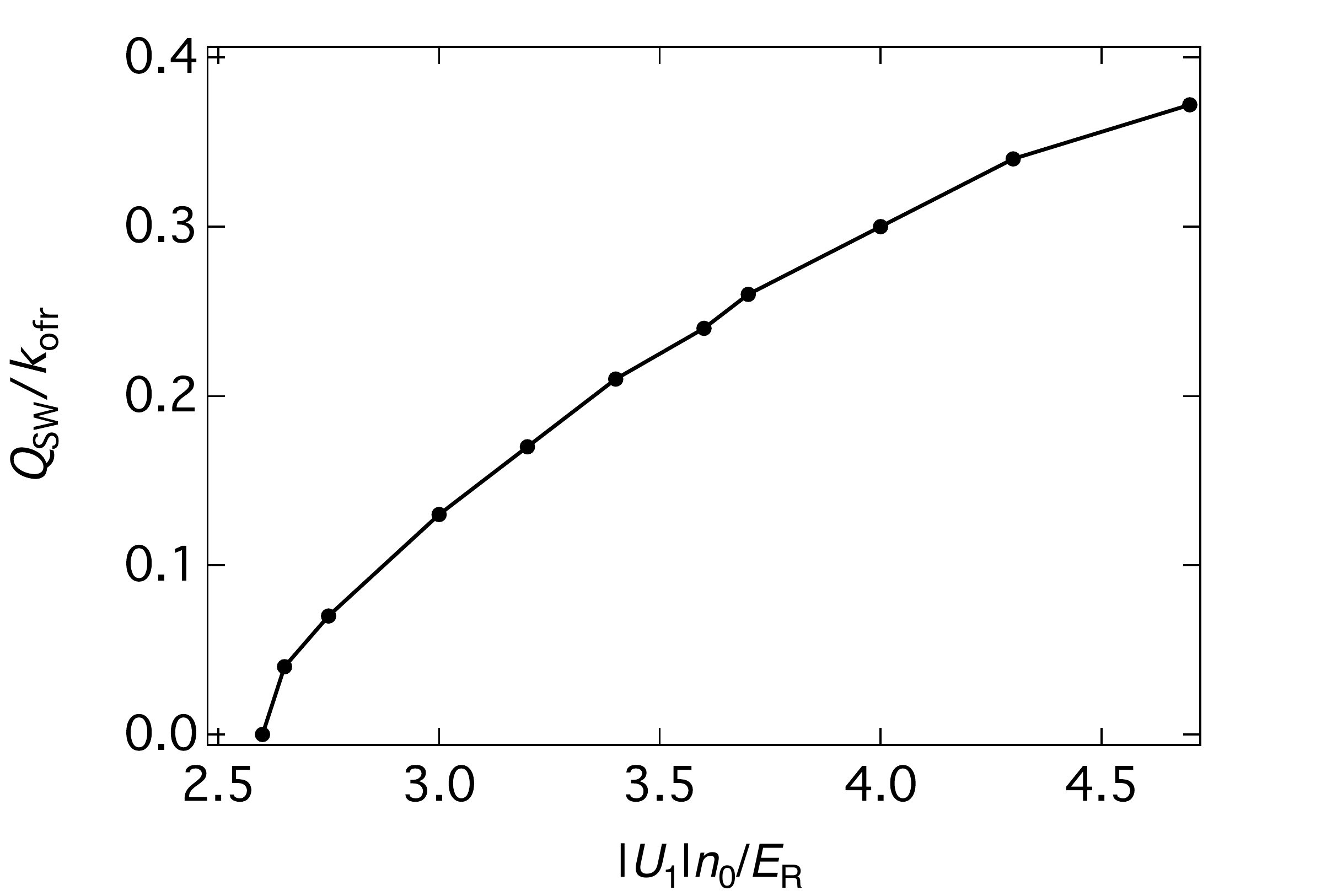}
\caption{Full-width of swallowtail by varying the intensity of the uniform
interaction $|U_1|$. In this figure, $U_2 n_0/E_R = 0.5$.}
\label{fig:halfwidth}
\end{figure}

To provide a better understanding of the swallowtail structures in the current system as well as the evolution of the energy spectra, let us first examine the system in the weak $U_2$ limit, such that $U_2 n_0 \ll E_F$ (see Fig.~\ref{fig:swallowtail-1}(a)(c)(e)).  In this limit, the mean-field Hamiltonian Eq. (\ref{eqn:H-mf}) becomes spatially homogeneous to the zeroth order of $U_2$. The saddle-point solutions for the uniform order parameter $|\Delta(x)| = \Delta_0$ and the chemical potential can be determined from the gap and the number equations
\begin{eqnarray}
\frac{1}{U_1} &=& - \frac{1}{L} \sum_{k} \frac{1}{2 E_{k}} \tanh\left(\frac{\beta E_k}{2} \right),
\label{eqn:gap-eq}
\\
n &=& \frac{1}{L} \sum_{k } \left[ 1 - \frac{\xi_k}{E_k} \tanh\left(\frac{\beta E_k}{2} \right) \right],
\label{eqn:num-eq}
\end{eqnarray}
where $L$ is the quantization length, $\xi_k = \epsilon_k - \mu$ denotes the single particle
dispersion shifted by the chemical potential, $E_{k} = \sqrt{\xi_k^2 + \Delta^2}$
is the quasiparticle dispersion, and $n$ is the number density.
\begin{figure}
\includegraphics[width=8.5cm]{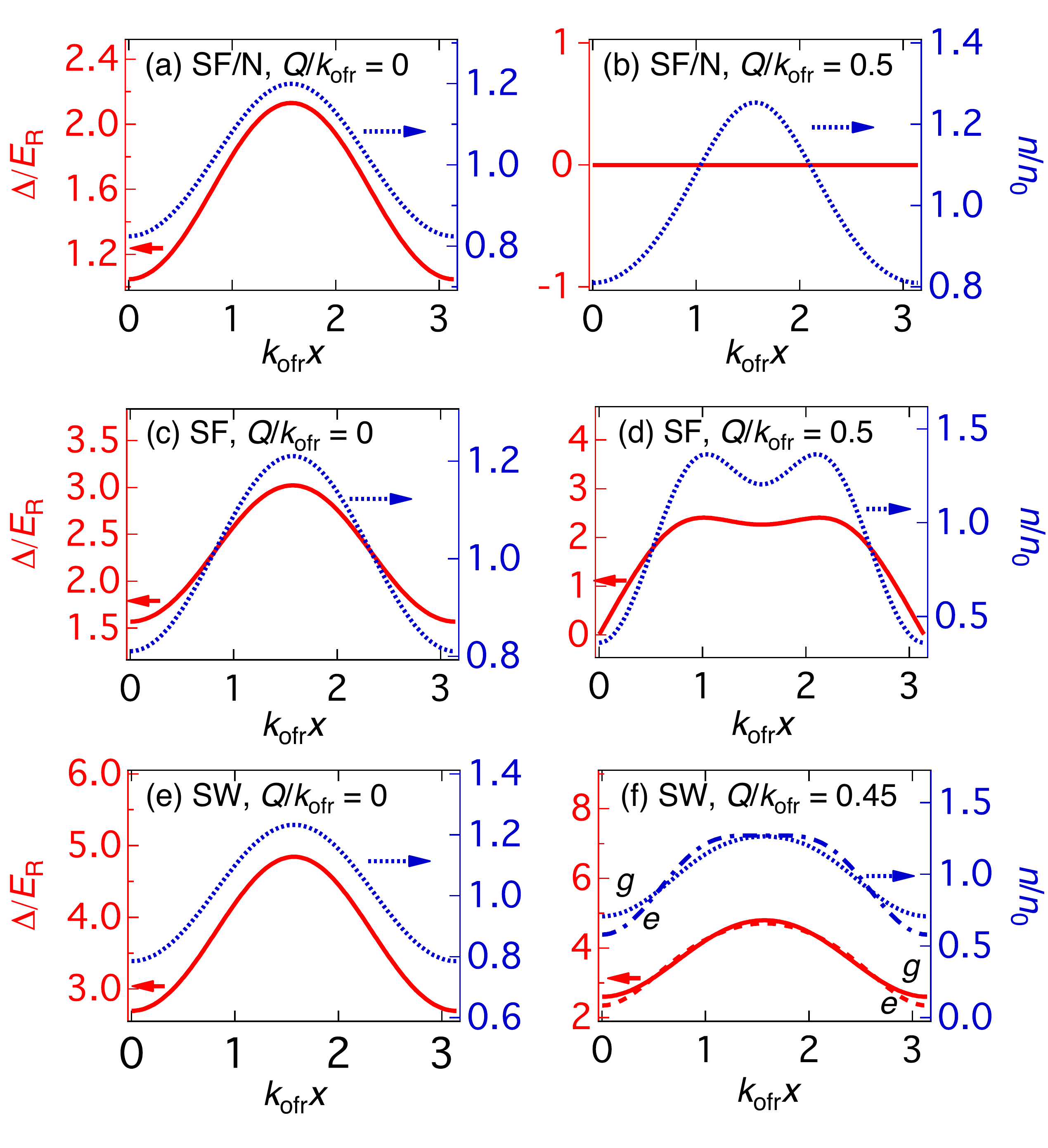}
\caption{(Color online) Spatial distributions of the order parameter (red)
and the particle number density (blue) within a unit cell. Three different cases
are plotted as: (a-b) SF/N with $|U_1| n_0/E_R = 2.0$;
(c-d) SF with $|U_1| n_0/E_R = 2.35$; and (e-f) SW with $|U_1| n_0/E_R = 3.0$.
Notice in (f) that two superfluid states as labeled by $g$ (ground) and $e$ (excited)
can be found, indicating the presence of the swallowtail structure. In this figure,
$U_2 n_0/E_R = 0.5$.}
\label{fig:spatial}
\end{figure}

In the absence of a supercurrent ($Q = 0$), the total energy of the superfluid phase is lower than the normal phase
for arbitrary attractive interactions ($U_1 < 0$). The condensation energy $E_C$, which is defined as the energy difference
between the normal state $E_N$ and the superfluid phase $E_S$, increases with $|U_1|$. By imposing a supercurrent with finite wave vector $Q$, the total energy of the superfluid acquires an addition kinetic term
\begin{eqnarray}
\label{eqn:E-super}
\delta E_S(Q) &=&  E_S(Q) - E_S(0)
\nonumber \\
&&
\hspace{-1cm}
=\frac{\hbar^2 Q^2}{2m} \sum_k \left[
\left(1 - \frac{\xi_k}{E_k}\tanh\left(\frac{\beta E_k}{2} \right)  \right)
- 2 Y_k  \epsilon_k
\right], \nonumber \\
\end{eqnarray}
where $Y_k = (\beta/2) {\rm sech}^2 (\beta E_k/2) $ is the Yoshida distribution.
The critical supercurrent wave vector is then defined as the critical $Q_c$ where the
gained condensation energy gets compensated by the kinetic energy associated
with the superfluid flow, i.e., $E_C = \delta E_S(Q_c)$. When the superfluid flows faster than the critical $Q_c$, the superfluid ground state undergoes a phase transition in the ground state to become normal.

The inclusion of a small interaction term $U_2$ will induce small spatially periodic modulations in
both the number-density and the order-parameter distributions on top of a uniform background. This gives rise to a periodic energy spectrum in the quasimomentum space. As a consequence, the energy of a superfluid flow with quasimomentum $Q$ also acquires a periodic structure
with multiple copies in higher Brillouin zones, with different copies intersecting at
positions of $Q = (\ell+1)k_{\rm ofr}/2$ ($\ell$ is an integer). For a weak uniform interaction
$|U_1|$, the condensation energy is smaller than the kinetic energy at the intersections,
and the supercurrent critical wave vector $Q_c$ is naturally smaller than $k_{\rm ofr}/2$,
as indicated by arrows in Fig.~\ref{fig:swallowtail-1}(a). In this case,
the system is a superfluid with conventional dispersion for slow supercurrent flow ($|Q|<Q_c$), and
becomes a normal fluid with fast flow ($|Q|>Q_c$). We denote such a case as SF/N.
On the other hand, when the uniform interaction $|U_1|$ becomes stronger, the condensation energy can be so large that the normal
state energy lies above $\delta E_S(k_{\rm ofr}/2)$. In this case, the swallowtail structure can emerge, under appropriate interaction parameters, as shown in Fig.~\ref{fig:swallowtail-1}(e).
\begin{figure}[t]
  \includegraphics[width=8.5cm]{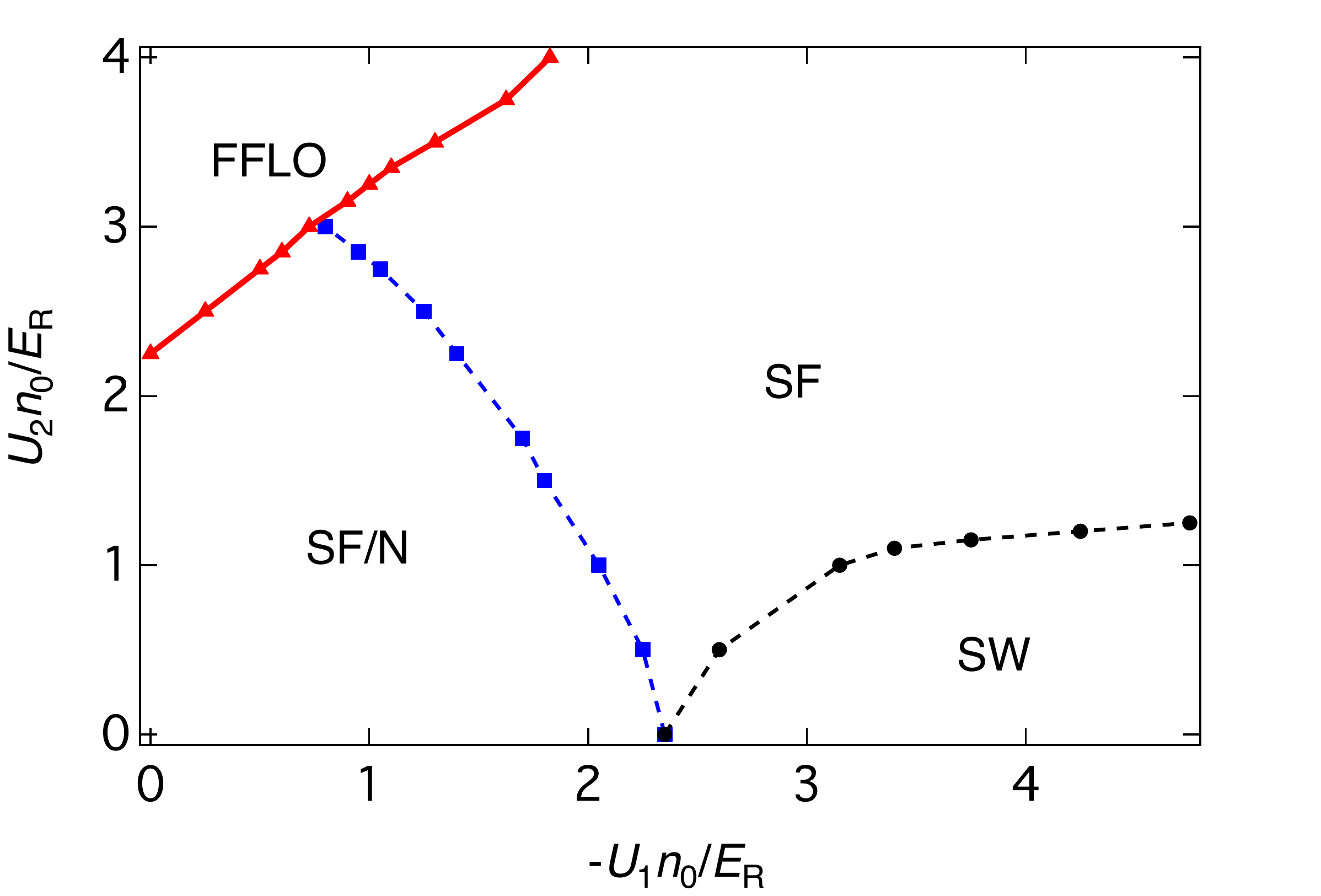}
  \caption{(Color online) Phase diagram for the parameter region $U_1 < 0$
  and $U_2 > 0$. Four regions discussed in texts can be identified on this diagram,
  and are separated by first-order (solid) or second-order (dashed)
  transition lines.}
  \label{fig:phase}
\end{figure}
\begin{figure}
\includegraphics[width=8.5cm]{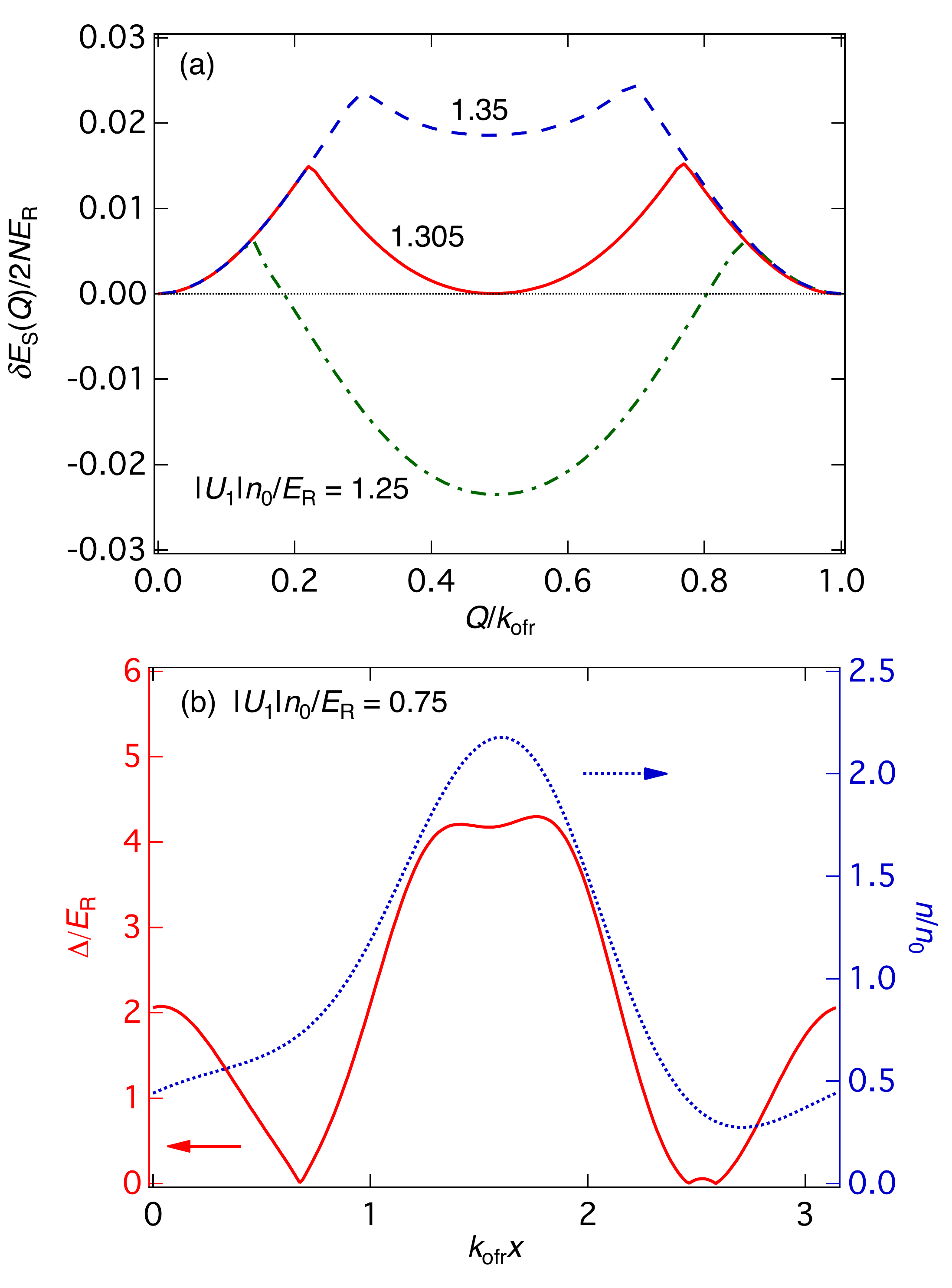}
\caption{(Color online) (a) Energy spectrum of the superfluid state with supercurrent
wave vector $Q$ for modulated interaction strength $U_2 n_0/E_R = 3.5$. Notice that
for small enough uniform interaction $|U_1|$, the superfluid state with $Q = k_{\rm ofr}/2$
can become energetically favorable, so that the ground state of the system becomes an
FFLO state. (b) Spatial distributions of the superfluid order parameter $\Delta (x)$ (red, solid)
and the particle number density $n(x)$ (blue, dotted) for the FFLO state within a unit
cell. Parameter used in this figure is $U_2 n_0/E_R = 3.5$.}
\label{fig:FF}
\end{figure}

As the modulated interaction strength $U_2$ increases, the spatial modulations
of number density and order parameter become more evident, and provide a stronger
periodic environment for the particles. The strong spatial periodicity of the system thus leads to 
an apparent band structure in the quasimomentum space with a gap opening at the boundaries 
of the Brillouin zone.
In other words, if we consider the superfluid phase with supercurrent wave vector $Q$ as
a macroscopic quantum state, the presence of an interaction term $U_2$ can induce an effective
coupling between supercurrents at different bands, such that a gap can be opened at
the location where the two bands intersect. In this case, if the uniform interaction $|U_1|$ is weak
such that the normal state energy lies within the lowest band, the system is in SF/N state as we have
discussed previously (see Fig.~\ref{fig:swallowtail-1}(b)). Conversely, if the uniform interaction is so large
that the normal state energy lies above the lowest band,
the ground state of the system is always a superfluid (SF) (see Fig.~\ref{fig:swallowtail-1}(d)).
If $|U_1|$ further increases, the uniform interaction overcomes the effect induced by $U_2$ and
restores the quadratic form of ground state energy. As a consequence, the system acquires
the swallowtail structure for $Q$ around $k_{\rm ofr}/2$, as indicated in Fig.~\ref{fig:swallowtail-1}(f).
The width of the swallowtail $Q_{\rm SW}$ for a fixed $U_2$ is shown in Fig.~\ref{fig:halfwidth},
from which one can see clearly that the swallowtail extends with increasing $|U_1|$. This
observation is consistent with the understanding that the uniform interaction favors swallowtail structure.

The real-space distributions of number density and order parameter for the three different cases
are listed in Fig.~\ref{fig:spatial}, using the same parameters as in Fig.~\ref{fig:swallowtail-1}(b)(d)(f).
Notice in Fig.~\ref{fig:spatial}(d) that the spatial distributions show clear double-peak structures within a unit cell for the case of
$Q = k_{\rm ofr}/2$ within the SF case. This is a clear signature of the higher band effect induced
by the interaction term $U_2$ at the gap-opening position.

\section{Phase diagram and the FFLO-like state}
\label{sec:pd}

With these understandings, we map out the phase diagram for the parameter region $U_1<0$ and $U_2 >0$.
As shown in Fig.~\ref{fig:phase}, the swallowtail structure can be found in a sizable parameter region (SW). Additionally, the SF/N and the SF regions can also be identified, consistent with our previous analysis. At small $|U_1|$ and large $U_2$ however, an FFLO-like state with a finite center-of-mass wave vector $Q = k_{\rm ofr}/2$ emerges on the phase diagram. The existence of this FFLO-like state can be understood along the same line of the previous energy spectrum analysis.

Starting from the SF/N region, a decreasing $|U_1|$ leads to a weaker non-linear interaction,
or equivalently, a stronger modulation in space. As a consequence, the periodicity effect becomes
more evident, which enhances the effective coupling between the two bands, and leads to the gap opening
between the two superfluid states at $Q = k_{\rm ofr}/2$. The superfluid energy $E_S$ of the lowest band
at $Q = k_{\rm ofr}/2$ thus decreases, as depicted in Fig.~\ref{fig:FF}(a). As the energy $E_S (k_{\rm ofr}/2)$
becomes lower than the energy $E_S(0)$ of the superfluid state at rest, the system undergoes a first-order
phase transition to become a superfluid state with a spontaneously generated finite center-of-mass momentum,
i.e., an FFLO-like state~\cite{ff, lo}. The phase boundary of this FFLO-like state is
determined by the position at which $E_S (k_{\rm ofr}/2)$ and $E_S(0)$ are degenerate (solid line in Fig.~\ref{fig:FF}(a)).
In Fig.~\ref{fig:FF}(b), we show typical spatial distributions of the number density and order parameter for the FFLO-like state. Notice that due to the strong modulated interaction $U_2$, the order parameter acquires nodes within a unit cell. Besides, the two degenerate states with $Q = \pm k_{\rm ofr}/2$ spontaneously break inversion symmetry, as can be usually observed in a single component Fulde-Ferrell state~\cite{ff}.
As a consequence, while the SF region is separated from SF/N and SW by second order phase boundaries, the FFLO-SF and FFLO-SF/N transition lines are of first order because of the abrupt change of the ground state from $Q=0$ to $Q = k_{\rm ofr}/2$ at the phase boundary.

\section{Conclusions}
\label{sec:summary}
We investigate the superfluid states in a quasi-1D Fermi gas under a spatially
modulated interaction, which can be realized by an optically tuned Feshbach resonance.
By calculating the ground state energy of the system via a BdG approach,
we obtain the energy spectrum of a superfluid flowing with wave vector $Q$, and reveal
a swallowtail structure near the boundary of the Brillouin zone within the regime
of strong attractive background interactions and weak modulated interactions.
This is consistent with the physical picture that the swallowtail structure is present as the effect of
a nonlinear interaction overcomes that induced by the periodic environment.
We map out the phase diagram by varying the interaction, and identify four different
regions therein, including an interesting FFLO-like region. For each region, we investigate typical spatial distributions of the order parameter
and particle density, which are helpful for the experimental detection.

\acknowledgments
This work is supported by NKBRP (2013CB922000), NFRP (2011CB921200, 2011CBA00200), NNSF (60921091),
NSFC (11274009, 11434001, 11374283), the Research Funds of Renmin University of China (10XNL016),
and the Program of State Key Laboratory of Quantum Optics and Quantum Optics Devices (KF201404). WY also acknowledges support from the ``Strategic Priority Research Program(B)'' of the Chinese Academy of Sciences, Grant No. XDB01030200.

\end{document}